\documentclass[%
 reprint,
 amsmath,amssymb,
 aps,
]{revtex4-2}

\usepackage{graphicx}
\usepackage{dcolumn}
\usepackage{bm}
\usepackage{comment}
\usepackage{nicefrac}
\usepackage[outercaption]{sidecap}  

\begin{document}

\preprint{APS/123-QED}

\title{Viscoplastic sessile drop coalescence}

\author{Vanessa R. Kern}
    \email{vanesske@math.uio.no}

\author{Torstein Sæter}%

\author{Andreas Carlson}
    \email{acarlson@math.uio.no}
\affiliation{%
  Department of Mathematics, Mechanics Division, University of Oslo, Oslo 0316, Norway
}%

\date{\today}

\begin{abstract}
The evolution of the liquid bridge formed between two coalescing sessile yield-stress drops is studied experimentally. 
We find that the height of the bridge evolves similar to a viscous Newtonian fluid, $h_0\sim t$, before arresting at long time prior to minimizing its liquid/gas interfacial energy. We numerically solve for the final arrested profile shape and find it depends on the fluid's yield stress $\tau_y$ and coalescence angle $\alpha$, represented by the Bingham number $\tau_y h_{drop} / \sigma$ modified by the drop's height-width aspect ratio. 
We present a scaling argument for the bridge's temporal evolution using the length scale found from an analysis of the arrested shape as well as from the similarity solution derived for the bridge's evolution.
\end{abstract}

\maketitle

Yield-stress fluids only flow when subject to a stress greater than their yield stress $\tau_y$. We rely on yield stress fluids. They allow us to squeeze toothpaste from tubes, lay mortar for bricks, plaster walls and even paint.
Much focus has been spent on yield-stress flows, with a growing interest in capillary flows such as surface spreading for inkjet printing and spray coating \cite{jalaal2021spreading, martouzet2021dynamic, jalaal2018gel, jalaal2019capillary, deRuiter2018drop},
viscous fingering \cite{lindner2000viscous},
and the adhesion of pastes \cite{barral2010adhesion}.

\begin{figure}[t]
\includegraphics[width=1\linewidth]{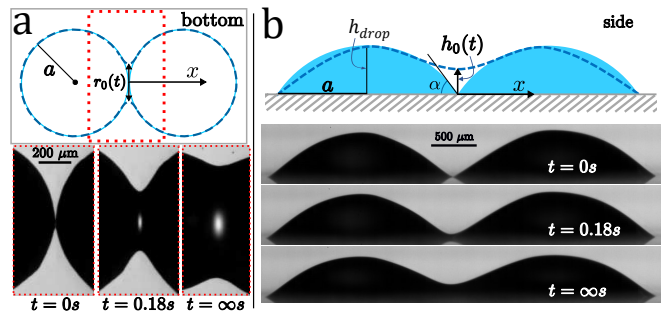}
\caption{\label{fig:1expsetup} 
Bottom (\textbf{a}) and side (\textbf{b}) view viscoplastic sessile drop coalescence with yield stress $\tau_y=27.5$ Pa. High speed imagery captures the coalescence event. 
Post-coalescence (\textbf{a}) the contact line spreads characterized by $r_0$, and
(\textbf{b}) the bridge grows characterized by its minima $h_0$, before halting due to the drop's yield stress. The initial height of the drop $h_{drop}$ and its footprint $a$ are related through the coalescence angle $\alpha$. Bottom and side view time sequences are presented from two different experiments with similar initial conditions.}
\end{figure}

We focus here on the dynamics of sessile viscoplastic drop coalescence utilizing both bottom-view and side-view imaging as shown in Fig. \ref{fig:1expsetup}\textbf{a,b}. 
When two drops coalesce, between them is formed a liquid bridge.
Much emphasis has been placed on the evolution of this bridge for both free and sessile Newtonian drops \cite{menchaca2001coalescence,thoroddsen2005coalescence,duchemin2003inviscid,aarts2005hydrodynamics}.
For sessile drops, the growth of the bridge can be characterized by its width $r_0$ as viewed from below and its height $h_0$ as viewed from the side.
The presence of a symmetry breaking substrate adds additional stress near the growing bridge, where the coalescence angle $\alpha$ has been shown to drastically alter its temporal evolution \cite{eddi2013influence}.
Different coalescence regimes have been identified, where the bridge height for inertially dominated flows grows as $h_0 \sim t^{0.5}$ \cite{duchemin2003inviscid} and for viscous dominated flows as $h_0 \sim t$ \cite{eggers1999coalescence,aarts2005hydrodynamics}, with corrections in the early time as $h_0 \sim t \ln{t}$ \cite{eggers1999coalescence}. For viscous flows, it has been shown $r_0 \sim t^{0.5}$ \cite{ristenpart2006coalescence} before transitioning to Tanner's-like spreading at long time \cite{bonn2009wetting,tanner1979spreading}. Inertial to viscous crossover depends on liquid properties and drop size \cite{paulsen2011viscous}.

Despite relevance in many motivating applications \cite{carrier2003coalescence,tcholakova2006coalescence,mahdi2008characterization}, far less is known about non-Newtonian coalescence.
For polymeric fluids, recently the effects of elasticity have been considered, showing polymer stretching affects spatial self-similarity, but not temporal growth
\cite{dekker2022elasticity}.
Other studies have shown new scaling laws for temporal bridge growth \cite{varma2020universality}, transitions from inertial to visco-elastic regimes \cite{varma2021coalescence} and effects of approach velocity during sessile coalescence \cite{sivasankar2021coalescence}.
When two yield-stress drops coalesce, their interface arrests before minimizing their surface energy. Arrested coalescence events have been previously observed for ``jammed'' interfaces in pickering emulsions \cite{pawar2011arrested} as well as for anisotropic colloids  \cite{pawar2012arrested,kraft2009self,sacanna2011shape}.

Here we find that despite the symmetry-breaking substrate, the shear-thinning properties of the drops considered, as well as a yield-stress fluid's ability to have yielded, liquid-like and non-yielded, solid-like, flow regions, the evolution of the liquid bridge evolution shares many similarities with the viscous Newtonian case in its growth as $h_0 \sim t$, its temporal collapse and its spatial self similarity, with the effect of the yield stress evident only in its final arrested shape.

\textit{Experiment} - Side and bottom view sessile coalescence events were imaged using a Photron Fastcam SA5 with a Nikon 200mm f/4 AF-D Macro lens and two teleconverters (Nikon 2x and 1.7x) resulting in resolutions of $\sim 6 \mu$m/pixel. Further processing in Matlab using partial area subpixel edge detection techniques afforded resolutions below $6 \mu$m/pixel \cite{trujillo2013accurate,trujillo2019code}.
Coalescence substrates were glass microscope slides (VWR Ca. 631-1550) rinsed with isopropanol (VWR Ca. 20922.364), plasma cleaned and kept in DI water until use. Immediately prior to use slides were blown dry with oil free compressed air. 
Coalescing drops were Carbopol 940 solutions with yield stresses varying from $\tau_y =2$ Pa to $50$ Pa measured using the stress growth test, details in the Supplementary Material \cite{supplementaryInfo}. Drop densities were assumed to be close to water $\rho \approx 1000$ kg/m$^3$ and liquid/gas surface tensions were assumed to be $\sigma = 0.053$ N/m as measured for a liquid bridge increasing in radius \cite{jorgensen2015yield}.

For consistency of comparison to viscous coalescence, only relatively spherical-cap shaped pre-coalescence sessile drops with aspect ratios $a/h_{drop} > 1.5$, and $\alpha < 45^\circ$ were considered.
To minimize deviations from the spherical-cap shape, drops with footprints $a < 1$ mm were preferred \cite{martouzet2021dynamic}.
When yield-stress drops spread on substrates, spreading arrests before drops minimize their surface energy as by the Young-Dupr\'e equation $\sigma \cos(\alpha) = \sigma_{SV}-\sigma_{SL}$ \cite{jalaal2021spreading}.
Therefore for our experiments, the smallest obtainable initial coalescence angles $\alpha$ we expect are $6.5^\circ$ to $34^\circ$, for $\tau_y = 2$ Pa and $50$ Pa, respectively \cite{supplementaryInfo,jalaal2021spreading,martouzet2021dynamic}.
Slides were plasma cleaned before each set of experiments to maintain a low substrate surface energy, thereby minimizing $\alpha$.
By knowing the initial drop size, approximate arrested angle $\alpha$ and footprint $a$ at which spreading will cease,
drops were simultaneously dispensed at known relative distances achieving both minimal $\alpha$ as well as contact line velocities below the initial bridge height velocity $\mathrm{d}h_0/\mathrm{d}t$.
Post coalescence, drops arrested within seconds rendering negligible evaporative losses.

\textit{Results} - We frame our problem as a 2D viscous flow parallel to the bridge justified by the small relative aspect ratio $h_0/r_0 \ll 1$, and low film Reynolds number \cite{duchemin2003inviscid,eggers1999coalescence,supplementaryInfo}.
As the fluid has a yield stress $\tau_y$, the system can be described using viscoplastic lubrication theory \cite{jalaal2021spreading,balmforth2019viscoplastic},
\begin{equation}\label{eqn:ViscoPlasticLubUnscaled}
\begin{split}
    \frac{\partial h}{\partial t} &+\frac{1}{6\eta}\frac{\partial }{\partial x}\left((3h-Y)Y^2\left|\frac{\partial P}{\partial x}\right|\right)= 0 \text{ and  }\\
    Y &=\max{\left(0,h-\tau_y \left|\frac{\partial P}{\partial x}\right|^{-1}\right)}
\end{split}
\end{equation}
with $h$, $P$, and $Y$ all functions of $(x,t)$. Here a minimal rheological model, the Bingham model, is used to represent Carbopol (see the Supplementary Material), where $Y(x,t)$ represents the boundary between the yielded and arrested regions of the drop, $h(x,t)$ the liquid/gas interface, $\eta$ the effective viscosity of the fluid, and $P(x,t)$ the capillary pressure. For lubrication theory to hold, one would expect that for the $\alpha$ at which our experiment is constrained that the lubrication approximation would be inaccurate, however this approximation has been shown to hold for $\alpha$ as large as $\alpha=67^\circ$ \cite{hernandez2012symmetric}.

\begin{figure}
\includegraphics[width=0.8\linewidth]{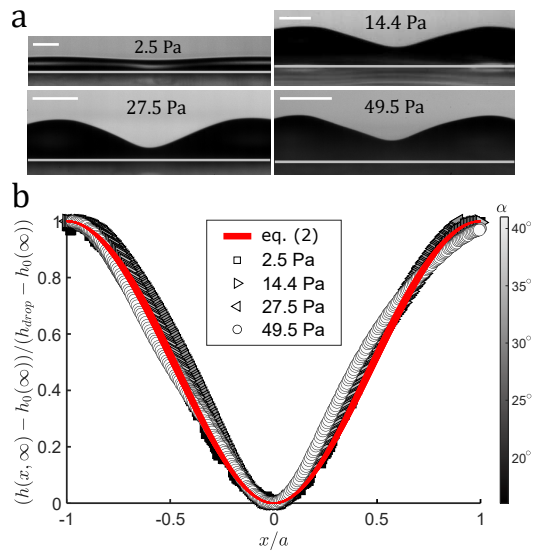}\centering
{\caption{(\textbf{a}) Post-coalescence ($t = \infty$) final arrested shapes for increasing yield stress $\tau_y$, each scale bar is $500 \mu$m. For similar coalescence angles $\alpha$, increasing $\tau_y$ decreases the arrested bridge height $h_f$. Increases in $\alpha$ serve to increase the arrested bridge height, as shown for the $49.5$ Pa case. (\textbf{b}) Comparison of the experimentally measured liquid/gas interfacial profiles $h(x,\infty)$ (open circles) against $x/a$ with their corresponding numerical solutions (solid lines) obtained from a scaled (\ref{eqn:integrate}), $\hat{h}\hat{h}_{\hat{x}\hat{x}\hat{x}}=J/(h_{drop}/a)^3$ with boundary conditions $\mathrm{d}\hat{h}/\mathrm{d}\hat{x}=0 \text{ at  } (\hat{x}=-1,0)  \text{ and } \hat{h}(\hat{x}=-1)=1$. Notably plotting a normalized $h(x,\infty)$ as shown collapses the numerically determined final profile heights for all 21 different experiments ranging $J =[0.01-0.47]$ and $h_{drop}/a = [0.15-0.58]$ onto a single curve.}\label{fig:2Arrested}}
\end{figure}

\textit{Final arrested shape} - Post Newtonian drop-drop coalescence, interfacial energies are minimized and the liquid/gas interface assumes a spherical-cap shape. However, post yield-stress drop-drop coalescence, the drop's bridge arrests at some fraction of its total height $h_f/h_{drop}$, see Fig. \ref{fig:2Arrested}\textbf{a}.

From (\ref{eqn:ViscoPlasticLubUnscaled}) we see that as $Y \rightarrow 0$ the yielded region of the drop vanishes. The shape of the arrested interface $h(x,\infty)$ can then be characterized by the non-linear ordinary differential equation obtained from (\ref{eqn:ViscoPlasticLubUnscaled}) in the limit $Y \rightarrow 0$,
\begin{equation}\label{eqn:integrate}
    h~\frac{\mathrm{d}^3 h}{\mathrm{d} x^3} = \frac{\tau_y}{\sigma}.
\end{equation}
Here the pressure $P$ is approximated as the local interfacial curvature $\sigma~ \mathrm{d}^2h/\mathrm{d} x^2$ with the small angle approximation and flow unidirectional towards the bridge $x=0$.
We numerically solve (\ref{eqn:integrate}) as a system of first order ODEs in python using SciPy's `solve\_bvp' function by prescribing the experimentally determined aspect ratio $h_{drop}/a$ and Bingham number $J = \tau_y h_{drop}/a$, scalings for the height $\hat{h}=h/h_{drop}$ and length $\hat{x}=x/a$ and boundary conditions $\mathrm{d}\hat{h}/\mathrm{d}\hat{x}=0 \text{ at  } (\hat{x}=-1,0)  \text{ and } \hat{h}(\hat{x}=-1)=1$,
representing symmetry at $h(-a,\infty) = h_{drop}$, symmetry at the bridge $h(0,\infty) = h_f$, and the height of the drop, respectively. 

\begin{figure*}
\includegraphics[width=\linewidth]{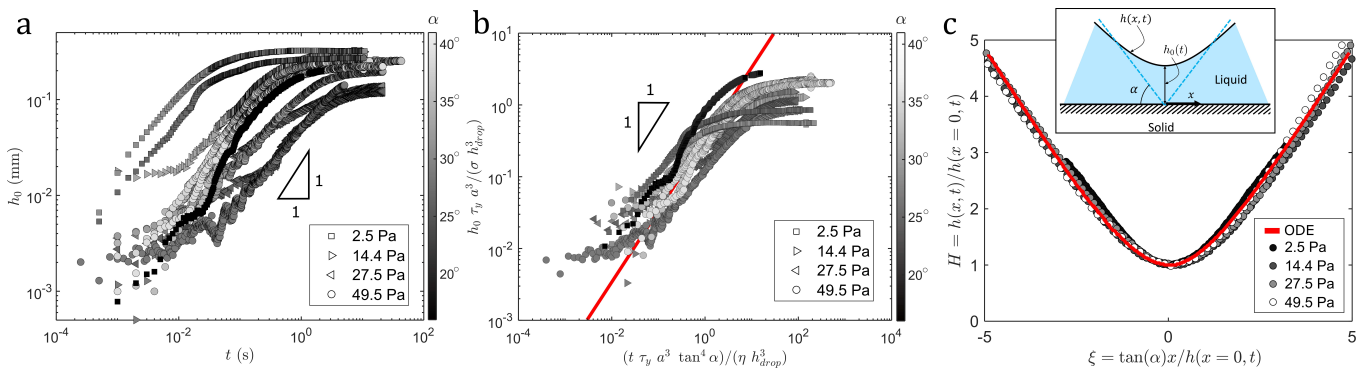}
\caption{\label{fig:4SelfSimilar2_DefSketch} 
Bridge evolution and self-similarity.
(\textbf{a}) Unscaled bridge $h_0$ evolution data.
(\textbf{b}) Scaled bridge $h_0$ evolution data. Inclusion of an effective viscosity $\eta$ in the x-axis scaling collapses the coalescence events. Unaccounted for variation in $\eta$ from varying $\alpha$ causes splay in the data. Here we use $\eta=[0.7,3.8,8.4,12.4]$ Pa$\cdot$s for the corresponding $\tau=[2,14.4,27.5,49.5]$ Pa.
(\textbf{c}) Scaled drop interface $H=h(x,t)/h_0(t)$ against the similarity variable $\xi$ in the linear regime of \textbf{b}. The solid line is the numerically determined similarity solution obtained when $Y\approx h$ from (\ref{eqn:ViscoPlasticLubUnscaled}) with scaling from (\ref{eqn:similarityVariables}), $\mathcal{H}-\xi\mathcal{H}'+1/V (\mathcal{H}^3\mathcal{H}''')'=0$ with one unknown parameter $V$ and boundary conditions $\mathcal{H}(0)=1$, $\mathcal{H}'(0)=\mathcal{H}'''(0)=0$, $\mathcal{H}''(\infty)=0$ and $\mathcal{H}'(\infty)=1$ \cite{hernandez2012symmetric}, $()'=\mathrm{d}/\mathrm{d}\xi$. Excellent agreement can be seen despite increasing $\tau_y$.
}
\end{figure*}

Fig. \ref{fig:2Arrested}\textbf{b} shows the results of numerically integrating a scaled (\ref{eqn:integrate}) compared with experimental data. We see good height profile agreement when the pre-coalesced drop shape is relatively spherical, with deviations a byproduct of initial deposition. We find in general that initial deviations from a spherical-cap shape cause deviations in the final arrested profile. Normalizing by the minimum bridge height as shown serves to collapse the numerically determined final profile heights for 21 different experiments ranging $J =[0.01,0.47]$ and $h_{drop}/a = [0.15,0.58]$ onto a single curve. More details about this numerical solution as well as a comparison of numerically recovered versus experimentally determined final bridge heights $h_f$ can be found in the Supplementary Material \cite{supplementaryInfo}. 

From the scaling of (\ref{eqn:integrate}) we can see that a non-dimensional quantity in the form of the Bingham number modified by the aspect ratio $J/(h_{drop}/a)^3$ arises, where a decrease in this quantity leads to an increase in the final arrested bridge height, recovering the intuition of a simple pressure balance shown in the Supplementary Material \cite{supplementaryInfo}.

\textit{Temporal evolution} - 
In Fig. \ref{fig:4SelfSimilar2_DefSketch}\textbf{a} 
we use the bridge height $h_0$ as a measure of the liquid bridge's temporal evolution and observe that $\alpha$, $\tau_y$ and $\eta$ significantly affect the coalescence velocity and arrested bridge height.
We observe for similar yield stresses $\tau_y$ that increasing $\alpha$ decreases the overall coalescence time and increases the final arrested bridge height $h_f$ \cite{supplementaryInfo}. We also observe for similar coalescence angle $\alpha$ increasing the yield stress $\tau_y$ decreases the bridge velocity and the final arrested bridge height $h_f$. 

Despite scanning a large parameter space, we observe the majority of the bridge's temporal evolution is similar to the viscous Newtonian case $h_0 \sim t$.
Similarities also appear to the Newtonian case in the radial direction for the drop's advancing contact line as $r_0 \sim t^{0.5}$ before transitioning to a long-time relaxation as $r_0\sim t^{0.1}$ similar to Tanner's law spreading \cite{tanner1979spreading}, shown in the Supplementary Material \cite{supplementaryInfo}.
From (\ref{eqn:ViscoPlasticLubUnscaled}) we can see that as $\tau_y \rightarrow 0$ or $\left|\partial P/\partial x\right| \gg \tau_y$, $Y \approx h$, reducing
(\ref{eqn:ViscoPlasticLubUnscaled}) to the 2D Newtonian lubrication equation \cite{hernandez2012symmetric,oron1997long}. This implies that the observed linear regime is a result of a rapid change in interfacial curvature such as in the vicinity of $h_0$.

Due to similarities in the temporal evolution of the liquid bridge with the viscous Newtonian case for the majority of the dynamics, 
we are then tempted to seek a similar universal re-scaling and similarity solution as can be shown for the viscous Newtonian case \cite{hernandez2012symmetric}. 
Following the logic of \cite{hernandez2012symmetric} when $Y\approx h$ we find similarity variables for (\ref{eqn:ViscoPlasticLubUnscaled}) of the form,
\begin{equation}\label{eqn:similarityVariables}
\begin{split}
    \mathcal{H}(\xi)&=\frac{h(x,t)}{vt}=\frac{h(x,t)}{h_0}, \text{ with} \\
    \xi&=\frac{\tan(\alpha)~ x}{vt}=\frac{\tan(\alpha)~ x}{h_0}.
\end{split}
\end{equation}
Here $v$ represents the coalescence velocity and $\tan({\alpha})$ is not simplified to $\alpha$ in order to account for the larger coalescence angles.
A scaling for the coalescence time $\hat{t}$ is also found to be of the form
\begin{equation}\label{eqn:timescaleevolve}
    \hat{t} \sim \frac{ a^3 \tau_y \tan^4(\alpha)~t}{\eta h_{drop}^3}.
\end{equation}
Here we scale by the characteristic length scale $(h_{drop}/a)^3 ~\sigma/\tau_y$ resulting from the dimensionless quantity $J/(h_{drop}/a)^3$ found previously, as well as a coalescence velocity $v \sim \sigma \tan^4(\alpha) / \eta$.
From (\ref{eqn:timescaleevolve}) an immediate concern arises in the handling of $\eta$, as $\eta$ is a function of the shear rate $\dot{\gamma}$ and consequently a function of time.
Without a good estimate of the yielded region in the vicinity of $h_0$, the shear rate $\dot{\gamma}$ cannot easily be estimated.
Instead we observe that scaling by $\hat{t}$ with all $\eta = 1$ Pa$\cdot$s, as shown in the Supplementary Material \cite{supplementaryInfo}, groups the data by $\tau_y$, with increases in $\tau_y$ decreasing the coalescence velocity. From the rheological data it can be seen that increases in $\tau_y$ increase the viscosity $\eta$ at every shear rate \cite{supplementaryInfo}, thus it can be expected that increases in $\tau_y$ should lead to decreases in the coalescence velocity $v$.

Since we find that the experiments can be grouped by coalescence velocity $v$ as a function of $\tau_y$, and because a linear time dependence in the growth of $h_0$ is still observed for the majority of each coalescence, we are inclined to believe there is an effective viscosity $\eta$ for each $\tau_y$ that can be used to scale the data. 
By writing $\eta$ from the definition of the coalescence velocity as
\begin{equation}\label{eqn:viscosityFunction}
    \eta \sim \frac{\sigma \tan^4(\alpha)}{v},
\end{equation}
we can extract effective viscosities during the linear regime $h_0 \sim t$. We find that for similar $\alpha$ the relationship between these viscosities $\eta$ and the fluid's yield stress $\tau_y$ is linear.
From the rheological data, we can then infer that the effective shear rate near the evolving bridge is $O(1)s^{-1}$.

The results of rescaling the data using effective viscosities for each $\tau_y$ determined at the shear rate $\dot{\gamma} = 4 s^{-1}$ are shown in Fig. \ref{fig:4SelfSimilar2_DefSketch}\textbf{b}.
As not all experiments are perfectly spherical-capped shaped initially, error inherently exists in the determination of the drop's aspect ratio $h_{drop}/a$ as well as coalescence angle $\alpha$. Additional error exists in the determination of $\eta$ as the shear rate near $h_0$ is unknown. However despite these uncertainties, scaling in this fashion does group the data around a collapsed linear regime.

For times $\hat{t}$ in this collapsed region of Fig. \ref{fig:4SelfSimilar2_DefSketch}\textbf{b}, we expect that the interface $h(x,t)$ should also collapse onto a self-similar shape for all coalescence angles $\alpha$ and yield stresses $\tau_y$.
By rescaling our data with the similarity variables of (\ref{eqn:similarityVariables}) we find that in this linear regime for varying yield stress a self similar collapse of the liquid bridge evolution does occur as presented in Fig. \ref{fig:4SelfSimilar2_DefSketch}\textbf{c}. 
Comparing this collapse with the numerically calculated similarity solution from (\ref{eqn:ViscoPlasticLubUnscaled}) as $Y \approx h$ shows excellent agreement, implying that for this system liquid bridge evolutions across varying yield stresses are indeed self-similar and can be collapsed using viscous Newtonian lubrication theory \cite{oron1997long,hernandez2012symmetric}.

\textit{Discussion} - Here we present a first look at the sessile drop coalescence of viscoplastic drops. Surprisingly we find that the majority of the bridge's evolution follows viscous Newtonian scaling laws.
We also find that 2D Newtonian lubrication theory captures the self similarity in the bridge's evolution across multiple yield stresses $\tau_y$ and coalescence angles $\alpha$.
We postulate this reduction to the Newtonian case can be explained as the viscoplastic lubrication equations (\ref{eqn:ViscoPlasticLubUnscaled}) reduce to the Newtonian lubrication equations for regions of rapidly changing interfacial curvature such as in the vicinity of $h_0$.
We find that both the yield stress $\tau_y$ and coalescence angle $\alpha$ affect the fractional height $h_f/h_{drop}$ at which the evolving bridge arrests and solve numerically for the final profile shape.
Here only experiments where drops were reasonably spherical-cap shaped were considered. These conditions, however, only represent a small set of cases for the coalescence of sessile yield-stress drops that can easily adopt highly non-spherical cap shapes \cite{martouzet2021dynamic}. It would then be interesting for future study to include drops with highly non-uniform initial liquid/gas interfacial curvatures and potentially larger initial coalescence angles $\alpha$. 

\textit{Acknowledgments} -  The authors gratefully acknowledge the financial support of the Research Council of Norway through the program NANO2021, project number 301138.

\bibliography{Bib}

\clearpage

\section*{\label{appen:A}Supplementary Material}

\section{Rheology}

\subsection{Carbopol preparation procedure}
Carbopol 940 (CAS. 9003-01-4) was briefly mixed in DI water between 300-1000 rpm using a Silverson L5M high shear laboratory mixer until dissolved to avoid clumping. 1M NaOH was added to the dissolved solution in an 8:1 wt:wt\% ratio of 1M NaOH:Carbopol 940. The solution was then gently mechanically mixed using a Kenwood Chef XL Titanium for approximately 10 days. The amount of dissolved Carbopol was controlled to vary the yield stress, with $[0.8,1.1,1.5,2]$ g/L Carbopol/Water resulting in fluid yield stresses of $[2.5,14,28,50]$ Pa, respectively.

\subsection{Rheological data}

\begin{figure}
\includegraphics[width=\linewidth]{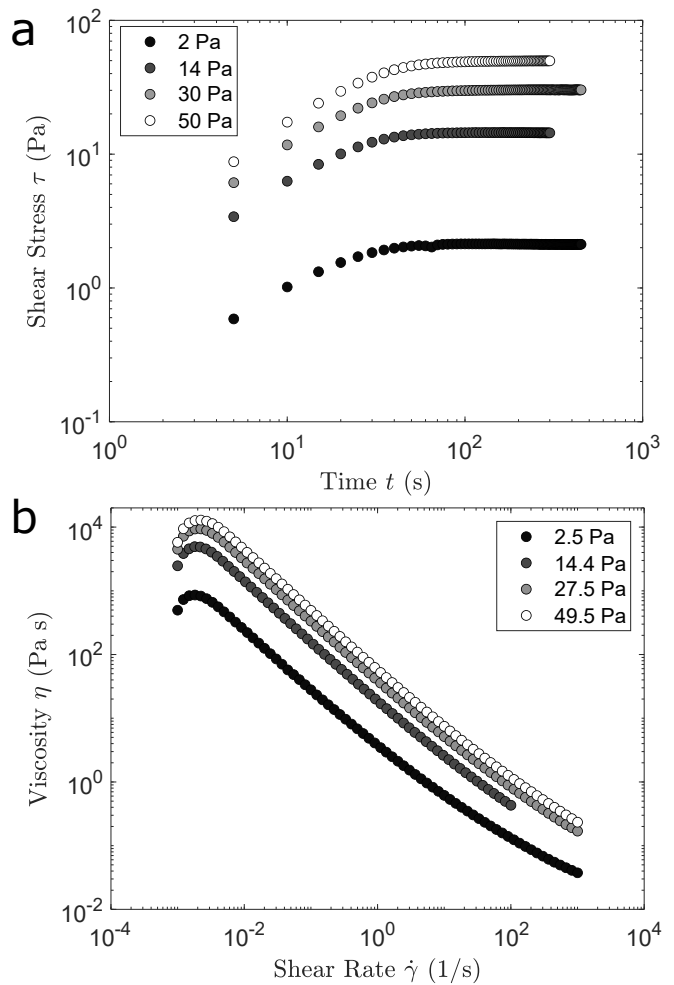}
\caption{\label{fig:CarbopolRheology} 
Rheological data for Carbopol 940.
\textbf{a} Measurement of yield stress using the constant oscillatory shear rate test. 
\textbf{b} Measurement of viscosity against shear rate.}
\end{figure}

No universal method for determining the yield stress of a material exists, with measured values varying depending on the type of test and the condition under which it was performed \cite{WPYieldStress}.
For instance, many complex fluids exhibit a reversible thixotropy, where at a given shear rate the viscosity of the fluid decreases over time, a behavior attributed to the breaking up of the fluid's internal structure \cite{ZOLEKTRYZNOWSKA201687}.
Measurements of the yield-stress behaviour are commonly carried out using a rotational rheometer, with some common and accepted methods including direct model fitting of measured stress versus strain data as by the Herschel-Bulkley model from (\ref{eqn:HBModel}), stress growth tests where a low shear rate, typically $0.01 s^{-1}$, is induced and the plateau in shear stress over time is taken to be the yield stress, or the oscillation amplitude sweep
with the yield stress accepted to lie within the between the initial drop in the loss modulus $G'$ and the crossover in the loss $G'$ and storage modulus $G''$ of the material.

Carbopol 940 solutions exhibit no thixotropic behavior, and for this study we chose the simplest and subjectively fastest of these yield stress measurement methods, the stress growth test with an oscillatory shear rate of $\dot{\gamma}=0.001s^{-1}$, shown in Fig. \ref{fig:CarbopolRheology}\textbf{a}.
Carbopol solutions were characterized on an Anton-Paar Rheometer with the ridged parallel plate geometry to prevent slip and before each set of experiments in order to remove any measurement error due to the degradation of the physical properties over time. 
Before each measurement a preshear step was included to consequently break and reheal the fluid's microstructure \cite{medhi2020investigation}.
It was observed that storing the solutions in plastic containers could lead to degradation of the yield stress by up to 30\% over the course of one month, therefore care was taken to store the prepared solutions in glass containers to avoid yield-stress degradation.

Carbopol 940 is known as a Herschel-Bulkley fluid with a stress $\tau$ versus shear rate $\dot{\gamma}$ dependency that can be expressed as
\begin{equation}\label{eqn:HBModel}
    \tau = \tau_y + K\dot{\gamma}^n \text{ for  } \tau\geq\tau_y,
\end{equation}
where $K$ and $n$ are known as the consistency index and power-law index, respectively (\cite{shafiei2018chemical},\cite{balmforth2000visco}). Flow occurs when $\tau > \tau_y$ with the fluid arrested when $\tau < \tau_y$ \cite{boujlel2013measuring}. Expressed in terms of viscosity,
\begin{equation}\label{eqn:viscosityshear}
    \eta = K \left|\dot{\gamma}\right|^{n-1} + \tau_y \left|\dot{\gamma}\right|^{-1}.
\end{equation}
We see that $n<1$ indicates a shear thinning flow and $n>1$ a shear thickening flow. For our experiments $n<1$, and increasing shear rates $\dot{\gamma}$ decrease the effective viscosity as shown in Fig. \ref{fig:CarbopolRheology}\textbf{b}.

\subsection{Deposition minimum $\alpha$}

\begin{figure}
\includegraphics[width=0.9\linewidth]{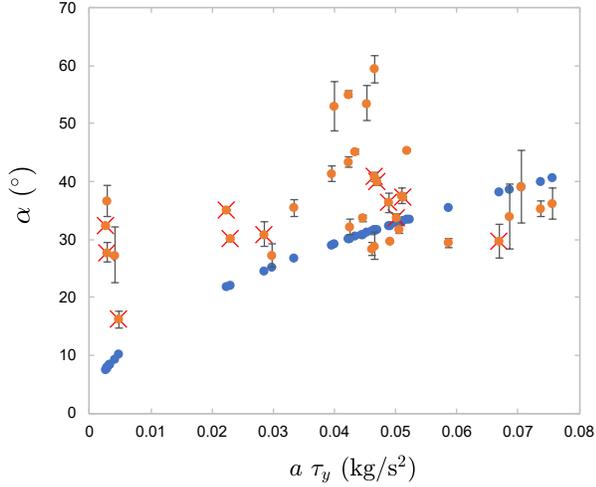}
\centering
\caption{\label{fig:MeasuredAnglesSI} 
Measured initial coalescence angles $\alpha$. Error bars represent the deviation between the two drops' coalescence angles. Red x's note the experiments considered, as they had minimal liquid/gas interfacial deviation from the spherical cap shape and minimal deviation in coalescence angles $\alpha$.
}
\end{figure}

When yield stress drops spread on substrates, spreading arrests before drops minimize their surface energy as by the Young-Dupr\'e equation $\sigma \cos(\alpha) = \sigma_{SV}-\sigma_{SL}$ \cite{jalaal2021spreading}. The contact angle at which spreading arrests can be predicted for our system by adapting (4) from \cite{martouzet2021dynamic} for a 0 surface energy solid as
\begin{equation}
    \alpha\approx\cos^{-1}\left({1-0.17\frac{\tau_y a}{\sigma}}\right).\label{eqn:alphaMin}
\end{equation}
A plot of the measured $\alpha$ against $\tau_y a$ is shown in Fig. \ref{fig:MeasuredAnglesSI}. 

\section{liquid bridge temporal evolution}

\begin{figure}
\includegraphics[width=\linewidth]{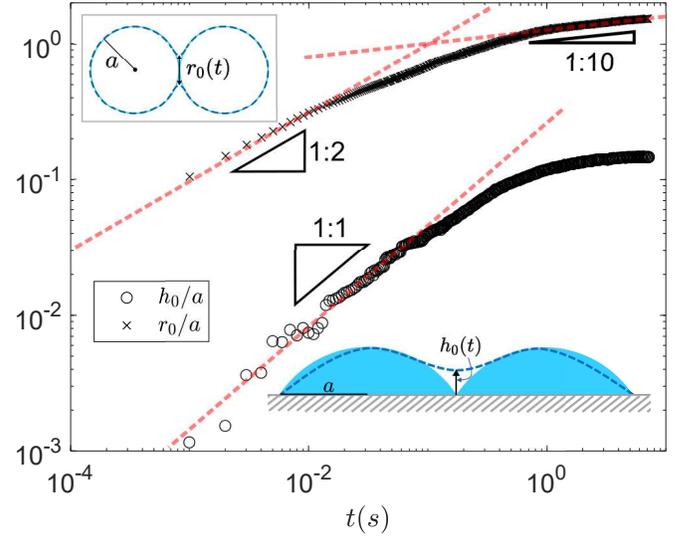}
\centering
\caption{\label{fig:2handr_evolv} 
Temporal evolution of the bridge height $h_0$ (circles) and radius $r_0$ (x's) scaled by the footprint $a$ for two different experiments with $\tau_y = 28$ Pa drops. At the early stages of coalescence $h_0$ is found to evolve linearly in time, similar to the viscous-dominated Newtonian case before arresting due to the liquid's yield stress. Similarly, the coalesced drops' contact line (CL) at $r_0$ evolves at early times as approximately $t^{0.5}$ before transitioning at long time to a slow relaxation, $t^{0.1}$.}
\end{figure}

As shown in Fig. \ref{fig:2handr_evolv} for the experiments of Fig. \ref{fig:1expsetup}, the temporal evolution of the coalesced drops' width $r_0$ and height $h_0$ are found to follow viscous scaling laws before arresting due to the drop's yield stress.

\begin{figure}
\includegraphics[width=\linewidth]{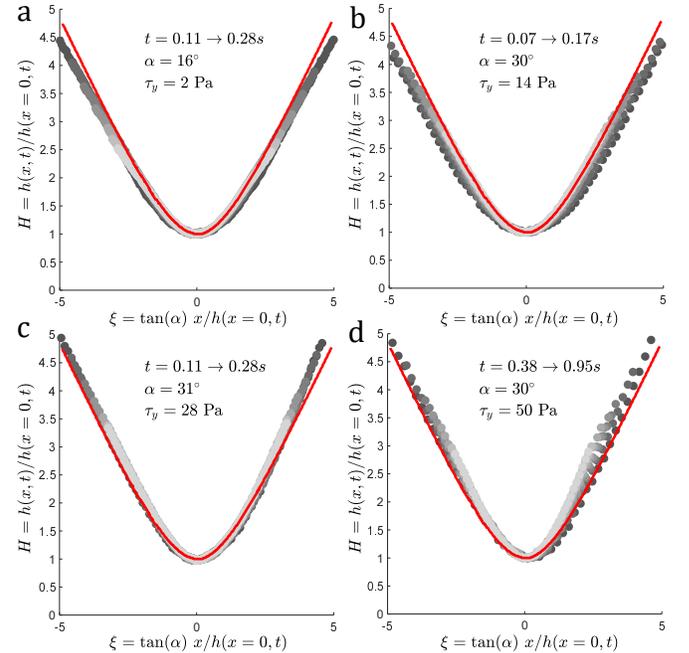}
\centering
\caption{\label{fig:selfsimOverTime} 
Bridge self similarity at varying times and yield stresses.
\textbf{a} $t = 0.11 \rightarrow 0.28$s and $\tau_y = 2$ Pa
\textbf{b} $t = 0.07 \rightarrow 0.17$s and $\tau_y = 14$ Pa
\textbf{c} $t = 0.11 \rightarrow 0.28$s and $\tau_y = 28$ Pa
\textbf{d}} $t = 0.38 \rightarrow 0.95$s and $\tau_y = 50$ Pa
\end{figure}

Additionally in the linear region of bridge evolution, it can be seen in Fig. \ref{fig:selfsimOverTime} that in the vicinity of $h_0$ the liquid bridge indeed grows self similarly at varying times as well as yield stresses.

\section{Equilibrium shape}

\subsection{Static approach}

An intuitive approach to predict the arrested drop's liquid/gas interfacial shape would have us assume $h_0$ arrests when the capillary pressure $\Delta P$ driving the coalescence no longer overcomes the yield stress $\tau_y$, as
\begin{equation}\label{eqn:PressureBalance}
    \tau_y = \Delta P = \frac{1}{2}\sigma \left(\frac{1}{r} - \frac{1}{R}\right)
\end{equation}
where $\Delta P$ is represented by the Laplace equation with $\sigma$ the surface tension, $1/r$ the curvature of $h(x,t)$ along $x$ and $1/R$ representing the negative curvature of the liquid/gas interface along $y$.

We can approximate the suction radius of curvature $R$ as,
\begin{equation}\label{eqn:Big_R}
    R \approx \frac{h_f}{1-\cos{\alpha}},
\end{equation}
by assuming the LG interface along the $y$-axis is a circular segment with a CL that spreads outward with an advancing contact angle equal to $\alpha$, and then assume $r$ has the form
\begin{equation}\label{eqn:Little_r1}
    r \approx \frac{h_0}{\tan{\alpha}} \frac{h_{drop}}{(h_{drop}-h_0)}+\epsilon,
\end{equation}
correct to some small constant $\epsilon$. Here $r$ is approximated as one half of the gap thickness between the original interface of the drop at the height $h_0$, corrected by the term  $h_{drop}/(h_{drop}-h_0)$ that goes to $\infty$ as $h_0\rightarrow h_{drop}$.

Substituting (\ref{eqn:Little_r1}) and (\ref{eqn:Big_R}) into (\ref{eqn:PressureBalance}) and rearranging we find
\begin{equation}\label{eqn:hfPredict}
    \frac{h_f}{h_{drop}} = \frac{\tan(\alpha)+\cos(\alpha)-1}{2\tau_y h_{drop}/\sigma+\tan(\alpha)} ,
\end{equation}
which can be approximated when $\alpha$ is small as
\begin{equation}\label{eqn:hfPredict_simp}
    \frac{h_f}{h_{drop}} = \frac{1}{J/\alpha+1} \text{ with  } J=\frac{\tau_y h_{drop}}{\sigma}.
\end{equation}
Here $J$ is known as the plasto-capillary number or Bingham number. 
For our experiments $\sigma$ and $h_{drop}$ remained relatively constant, therefore from (\ref{eqn:hfPredict_simp}) we can glean the intuition that the observed final arrested shape is a function of primarily the yield stress $\tau_y$ and also the coalescence angle $\alpha$, with increasing $\alpha$ or decreasing $\tau_y$ increasing $h_f/h_{drop}$. 

\subsection{Numeric Approach}

\begin{figure}[!t]
\includegraphics[width=0.8\linewidth]{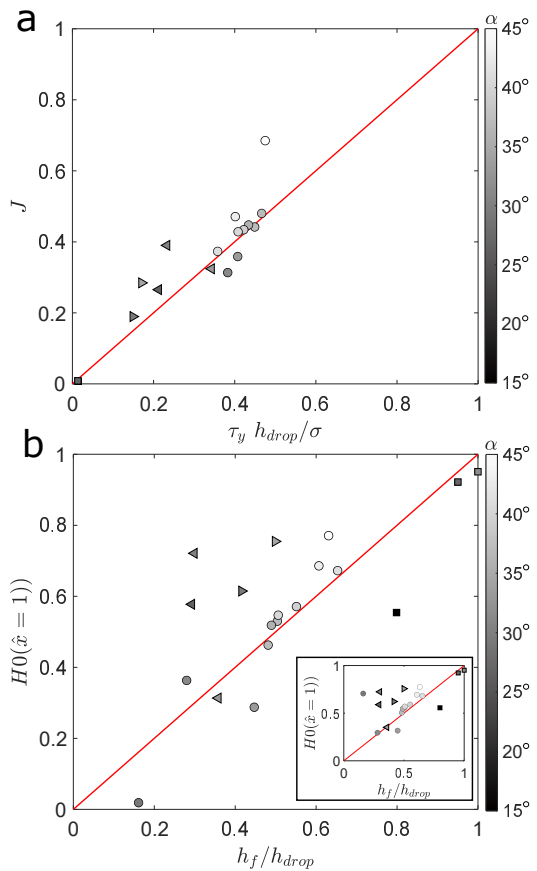}
\centering
\caption{\label{fig:MeasuredLambda} 
\textbf{a}. Numerically determined $J$ from (\ref{eqn:numInteg}) with boundary conditions (\ref{eqn:numIntegBCs}) against experimentally measured values. $J$ can be reasonably recovered from (\ref{eqn:hxxxsolved}) upon prescription of the final bridge height.
\textbf{b}. Numerically determined $h_f/h_{drop}$ from (\ref{eqn:hxxxsolved},\ref{eqn:numInteg}) with boundary conditions $H0(\hat{x}=0)=1$, $H1(\hat{x}=0,1)=0$  against experimentally measured values. The value of $h_f$ can be fairly well recovered for $J=[0.01,0.47]$ and $b=[0.15,0.58]$. The inset shows a comparison with the $h_f$ determined when using the small angle approximation.
}
\end{figure}

The pressure gradient $P_x$ from (\ref{eqn:ViscoPlasticLubUnscaled}) can be approximated as 
\begin{equation}\label{eqn:PressureGradient}
    P_x = \sigma\left( \frac{h_{xx}}{(1+h_x)^{3/2}}\right)_x,
\end{equation}
where $()_x = \mathrm{d}/\mathrm{d}x$.
Here the pressure is represented as the curvature in the x-direction without the small angle approximation. 
After applying scalings for the height $\hat{h}=h/h_{drop}$ and length $\hat{x}=x/a$, we can solve for $\hat{h}_{\hat{x}\hat{x}\hat{x}}$ obtaining 
\begin{equation}\label{eqn:hxxxsolved}
    \hat{h}_{\hat{x}\hat{x}\hat{x}} = \frac{J}{b^3} \frac{ \left( 1+b^2 {\hat{h}_{\hat{x}}}^2 \right)^{3/2}} {\hat{h}} + 3 b^2 \frac{ h_{\hat{x}} {h_{\hat{x}\hat{x}}}^2 } {\left( 1+b^2 {\hat{h}_{\hat{x}}}^2\right)},
\end{equation}
where here $b$ represents the aspect ratio $h_{drop}/a$. It can be seen that when $b<<1$, only the yield stress dependent term remains 
\begin{equation}\label{eqn:integrateScaled}
    \hat{h}\hat{h}_{\hat{x}\hat{x}\hat{x}}= J/b^3,
\end{equation}
In the main text, we solve (\ref{eqn:integrateScaled}) to determine the shape of the final bridge height, however a solution from this governing equation whilst capturing the shape of the final profile can be made more accurate in its prediction of the nominal value of $h_f$ by solving without any small angle approximation.

We approach a solution of (\ref{eqn:hxxxsolved}) in two ways. First we solve for $J$ using SciPy's `solve\_bvp' to solve the following system of 3 ODEs and one free parameter $J$
\begin{equation}\label{eqn:numInteg}
\begin{split}
    H_0' &= H_1, \\
    H_1' &= H_2, ~~\text{and} \\
    H_2' &= \left| \frac{J  (1+b^2 H_1^2)^{1.5}}
    {( b^3 H_0 )} + \frac{ 3 b^2 H_2^2 H_1 }
    {(  (1+b^2 H_1^2))} \right|,
\end{split}
\end{equation}
with 4 boundary conditions
\begin{equation}\label{eqn:numIntegBCs}
\begin{split}
    H0(\hat{x} = 0) &= 1, \\
    H0(\hat{x} = 1) &= h_f/h_{drop}, ~~\text{and} \\
    H1(\hat{x} = 0,1) &= 0.
\end{split}
\end{equation}
The results of solving for $J$ are shown in Fig. \ref{fig:MeasuredLambda}\textbf{a}, where good agreement can be seen invariant of yield stress $\tau_y$ or coalescence angle $\alpha \approx b$. 

\begin{figure}[t!]
\includegraphics[width=0.9\linewidth]{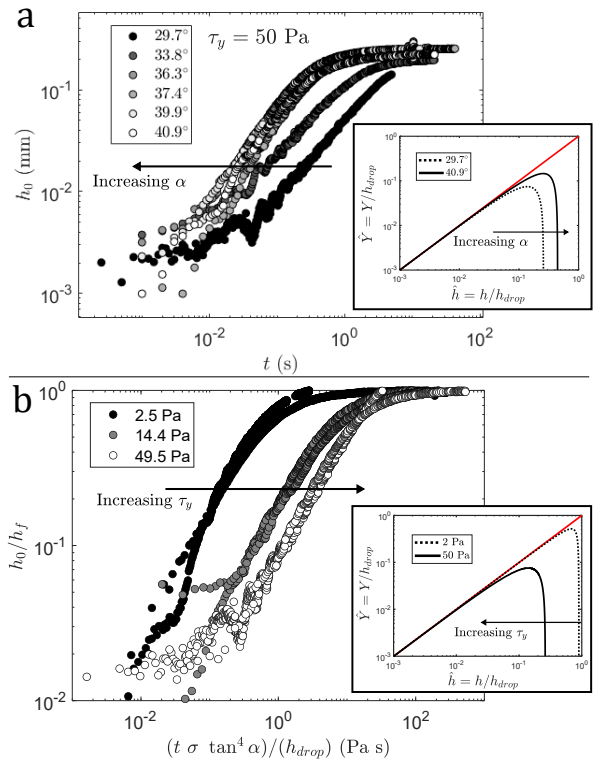}
\caption{\label{fig:3evolutions} 
Temporal bridge evolutions.
(\textbf{a}) For yield stress $\tau_y = 50$ Pa, increasing the coalescence angle $\alpha$ decreases the coalescence velocity and increases the arrested bridge height, however the evolution remains approximately $h_0 \sim t$. The inset shows the relevant phase portrait $\hat{Y}$ against $\hat{h}$ from (\ref{eqn:ScaledY}) with $h_{drop}=1$ mm. It can be seen that for the bulk of the evolution $Y\approx h$, thereby supporting the idea that $h_o \sim t$.
(\textbf{b}) Bridge evolutions grouped by yield stress $\tau_y = 50$ Pa. Increasing $\tau_y$ decreases the coalescence velocity.
The inset shows the relevant phase portrait $\hat{Y}$ against $\hat{h}$ with $h_{drop}=1$ mm and $\alpha=30^{\circ}$. Here also it can be seen $Y\approx h$ for the bulk of the evolution.
}
\end{figure}

We can also solve (\ref{eqn:numInteg}) for the final interface shape independent of any post-coalescence measured quantities by prescribing $J$ and removing the boundary condition $H_0(\hat{x}=1) = h_f/h_{drop}$. In Fig. \ref{fig:MeasuredLambda}\textbf{b} we see a plot of the numerically calculated final arrested bridge heights $h_f/h_{drop}$
found by solving these 3 ODEs. Here we see that the predictive capacity of (\ref{eqn:hxxxsolved}) versus (\ref{eqn:integrateScaled}) in determining the final arrested bridge height is increased through inclusion of the additional terms of (\ref{eqn:PressureGradient}).

\section{When is $Y\approx h$?}

When $Y \approx h$, the viscoplastic lubrication equations reduce to the Newtonian case. It would be useful to understand when during the bridge evolution $Y\approx h$.
Near $h_0$ we find $\left|\partial P/\partial x\right|$ can be written in terms of the coalescence angle $\alpha$ and the minimum bridge height $h_0$ as
\begin{equation}\label{eqn:dPdx}
    \left|\frac{\partial P}{\partial x}\right|
    \sim \sigma\frac{\tan^2(\alpha)}{h^2_0}\frac{(h_{drop}-h_0)}{h_{drop}},
\end{equation}
by using the approximation $P \approx \sigma/r$ as $R \rightarrow \infty$ with $\Delta x$ taken to be $h_0/\tan(\alpha)$.
From the viscoplastic lubrication equations $Y$ can then be rewritten as
\begin{equation}
    \hat{Y} = \hat{h} - \frac{J}{\tan^2({\alpha}) } \frac{\hat{h}^2 }{(1-\hat{h})}
    \label{eqn:ScaledY}
\end{equation}
where $\hat{Y}=Y/h_{drop}$, $\hat{h}=h_0/h_{drop}$ and $J$ is the Bingham number $\tau_y h_{drop}/\sigma$.
Solving for $\hat{h}$ when $Y=0$ recovers two equilibrium points, $\hat{h} = 0$ representing the beginning of the coalescence and $\hat{h} = 1/(J/\tan^2{\alpha}+1)$ representing the end of the coalescence. This second point is strikingly similar to (\ref{eqn:hfPredict_simp}), however a difference arises as we assume $R \rightarrow \infty$.

Fig. \ref{fig:3evolutions}\textbf{a} shows the results of plotting $\tau_y=50$ Pa experiments with varying $\alpha$. It can be seen that increasing $\alpha$ decreases the coalescence velocity, however the bridge evolution remains approximately $h_0 \sim t$ during the bulk of the evolution. The inset shows the results of plotting the phase portrait $\hat{Y}$ against $\hat{h}$ from (\ref{eqn:ScaledY}). It can be seen as $\alpha$ increases the final arrested height increases, but for the bulk of the evolution $Y$ remains $\approx h$. Fig. \ref{fig:3evolutions}\textbf{b} shows the effect of $\tau_y$ on the bridge evolution. Scaling time to recover a viscosity we can collapse the effect of $\alpha$ on the evolution and group the data by yield stress $\tau_y$. Increasing $\tau_y$ decreases the coalescence velocity as the effective viscosity increases, however similar to the varying $\alpha$ case we still recover a region where $h_0 \sim t$.

\end{document}